# Gender differences in research grant allocation - a mixed picture[1,2]

Peter van den Besselaar[3] & Charlie Mom[4]

## Abstract

Gender bias in grant allocation is a deviation from the principle that scientific merit should guide grant decisions. However, most studies on gender bias in grant allocation focus on gender differences in success rates, without including variables that measure merit. This study has two main contributions. Firstly, it includes several merit variables in the analysis. Secondly, it includes an analysis at the panel level where the selection process takes place, and this enables to study bias more in-depth at the process level. The findings are: (i) After controlling for merit, a consistent pattern of gender bias was found in the *scores*: women receive significant lower grades than men do. (ii) The scores are an input into the two-step *decision-making* process, and this study shows bias *against* women in the first selection decision where 75% of the applications are rejected, and bias *in favor* of women in the second (final) selection decision. (iii) At the level of individual *panels*, the analysis shows a mixed pattern of bias: in some panels the odds for women to receive a grant are lower than for men, whereas in other panels we find the opposite, next to panels with gender-neutral decision making. (iv) In the case under study, at an aggregated level the allocation of grants seems balanced. (v) The mixed pattern at panel level seems to relate characteristics such as the panel composition, and the level of gender stereotyping.

**Keywords**: gender; gender bias; disparities; discrimination; peer review; panel review; research grants; research funding; ERC; European Research Council.

[1] This paper benefitted from questions and comments of Laura Cruz-Castro (CSIC, Madrid), Naomi Ellemers (Utrecht University), Harry Ganzeboom (Vrije Universiteit Amsterdam), Torger Möller (DZHW, Berlin), Barbara Romanowicz (ERC Council, UC Berkeley), and Luis Sanz-Menendez (CSIC, Madrid). This is a result of the GendERC project, which received support from the ERC (grant 610706).

[2] Some minor changes compared to the earlier version.

[3] Professor emeritus, Vrije Universiteit Amsterdam, The Netherlands. Email: p.a.a.vanden.besselaar@vu.nl

[4] Email: charliesannemom@gmail.com

**Introduction**

Although gender differences in science are becoming smaller over the years, they have not disappeared [1, 2]: Women apply less often for grants [3], have slower careers [4, 5] have a lower chance to be hired as full professor [6], and earn less money[7]. These differences may be based on an average lower past performance of women [8, 9], but if not, they are a deviation from Merton's *universalism* norm, and are signs of gender *bias* [10, 11].

This paper addresses gender differences in grant allocation: do women have a lower probability to win a grant[5], and if so, is this due to merit related differences, or is it the effect of implicit or explicit bias? Several recent reviews suggest absence of gender bias in grant allocation [2, 12]. However, as most research has not included merit related variables, it is too early to draw such conclusions and indeed the omission of merit variables is seen as a core issue to be solved [13].

Selection of grant applications is generally done through peer review – often within panels. Panelists deploy a variety of criteria [14], of which productivity and impact are two important ones. There is ample evidence that women on average score lower on these two criteria than men do, and that men are particularly overrepresented among the top-performers [8, 9]. This observation could provide a merit-based explanation of gender disparities in grant success. But there is also evidence on the problems of peer review processes [15-19], as it is unable to select the best applicants [20-22], has a low predictive validity [23, 24], has a low inter-reviewer reliability [25-27], and is characterized by conservatism and risk-avoidance [28-29]. These problems should be expected, as small-group research has showed that decision-making is influenced by group characteristics and processes. So bias may be unavoidable [30, 31].

A core mechanism that influences group decision-making is *gender stereotyping*, that is panelists implicitly assuming men are better suited for science than women [32-35], which leads to differences in how men and women are assessed. Women obtain lower scores and have a lower probability to receive a grant than men with equal quality and performance [34]. Policies to counteract stereotyping have been implemented, such as more female panel members and training panel members about gender stereotypes. Whether these policies are effective is uncertain, as female panelist with a successful academic career may have internalized the same stereotypes (the *queen bee* phenomenon [36]). And training how to counteract gender stereotyping in grant panels has not been studied as far as we can see.[6] In this paper we investigate (i) the prevalence of gender bias in panel decisions, and (ii) whether differences bias can be explained by panel characteristics and processes.

---

[5] In theory there can also be gender bias against men, and as we will see later this does occur.

**Earlier research on bias in grant allocation**

Gender bias in grant peer review was put on the agenda by the Wennerås & Wold (W&W) study of the Swedish medical research council [37], a paper which has become a point of reference in the debate. In their highly cited[7] reviews [1, 2], Ceci and colleagues criticize the W&W paper on methodological grounds, confront it with several other studies, and then conclude that the "weight of evidence points overwhelmingly to a gender-fair grant review process. There are occasional small aberrations, sometimes favoring men and sometimes favoring women; all of the smaller-scale studies failed to replicate Wennerås and Wold's provocative findings, and all but one of the large-scale studies did as well – however this one study was reversed after a more ambitious joint reanalysis". [2, p3157]

This evidence still requires further inspection. One issue is that W&W focused on gender bias in the *scores*, whereas most of the studies criticizing W&W are comparisons of the *success rates*. E.g., the studies on the Canadian Research Council [38], and on the British MRC [39, 40] report the absence of gender differences in success rates only. A study on the Wellcome Trust did take past performance into account [39], as did a replication of the W&W study. The latter found equally strong nepotism, but no gender bias in the scores anymore [42].[8]

The larger studies covered in Ceci et al,'s review include a variety of covariates. A study of grant allocation at the NSF, the NIH and the DoA found about equal success rates for men and women, after controlling for age, academic degree, institution, grant type, institute, and application year [43]. For subsequent applications, male application rates were higher – but that refers to possible gender bias in the application phase, not in the selection phase. Another next study on the Australian Research Council found that women have equal success rates compared to men, but women are a rather small minority of the applicants [44, 45]. This holds when looking at the individual disciplines. The study did not include variables for past

---

[6] The *queen bee* mechanism [36] has been explained using meso level arguments (the organizational cultures in science are masculine) and micro level arguments (the negative individual experiences of women during the academic career).

[7] The reviews belong to the top 1% highest cited papers in their discipline (WoS).

[8] One study notes that despite equal success rates, female researchers are much less applying for grants [40]

performance or other forms of merit, which according to the same authors is "an important limitation in our research program" [13, p167]. The third study on the selection of doctoral and postdoctoral fellowships at a German research funder included variables like nationality, gender, field of study, and type of research organization, next to several merit indicators [46]. Some of these merit indicators are part of the selection process (e.g., the rating by reviewers and by the council staff) and cannot be used for evaluating bias. The remaining performance indicators were the final grade received for the university degree (for the predoctoral fellowships) and the number of publications (for the postdoctoral fellowships). The study found gender bias in the predoctoral, but not in the postdoctoral fellowships. A fourth study focused on the applicant-reviewer similarity within the NSF economics domain [47] and found that women fared well when rated by male reviewers, but less when rated by female reviewers, but did not report whether this leads to gender biased outcomes. The fifth study on NIH grants showed that in some of the six biomedical fields men have a higher success rate, and in other women do, but overall there was no difference [48].

A *meta-analysis* of 21 studies found an advantage of men above women of about 7% [49], but an improved meta-analysis including 66 studies found no significant differences between the success rate of men and women when controlling for discipline, country and grant type [13].

More recent studies tend to find gender bias in grant allocation, although not consistently. A Canadian study found that grants for women are substantially smaller than grants for men [49]. In a Dutch case study, gender differences in success rates were found favoring men [50]. This study used the scores the applicants received from the reviewers as quality measure, but in our view these are a *dependent* variable – and may be biased themselves as we will see in the case presented below. A Belgian study found no gender differences in success rates [51], and a natural experiment at a Canadian funding organization did find a gender gap, but also that funding instruments where the evaluation focused on the project and not on the PI, the gap was substantially smaller [52].

Summarizing, the large majority of studies about gender bias in grant allocation focus on differences in success rates,[9] with diverging results. Only a few studies did include some past performance or other merit variables. In most studies, the level of analysis is the funding organization or the funding instrument, and not the panel where the main decisions are made possibly influenced by group dynamics and gender stereotyping. The present study contributes by addressing these two issues: (i) It includes a series of merit indicators, and (ii) it studies bias at the panel level.

---

[9] Some studies emphasize gender differences in *grant application behavior* [40, 43, 53]. Compared to the share of women in the total qualified researcher population (the potential applicants), the share of women among grant applicants is still low [53] although it has improved over time [41]. That women do apply less (self-selection) can be choice in which case it cannot be classified as bias; but is can as well be the effect of gender inequality in the private sphere [54] or in the work sphere [8].

**The case**

We study the 2014 Starting Grant of the European Research Council (ERC), and obtained informed consent from 3,030 applicants (about 95%). This ERC grant is the most prestigious grant (1.5 million Euro for five years) in the EU for early career researchers, and it is expected to strongly contribute to career opportunities of those getting the grant [55-58]. Everyone can apply up to seven years after the PhD, but the application needs to mention a European Union based 'host institution'. Overall female applicants have a slightly lower success rates than men (11.2% versus 11.9%), but it differs strongly between the 25 disciplinary panels. The selection procedure consists of two steps. In the first, about 75% of the applicants are rejected. The remaining 25% proceed to the second step, where about half of them win the starting grant.

The panel processes are hardly formalized, neither are the criteria deployed by the panelists. The only criterion is excellence of the project and the investigator. Excellent investigators have shown creative independence thinking, the ability to do groundbreaking research, and have moved beyond the state of the art. Finally, panels consist of excellent researchers in their respective fields and should therefore decide among themselves what excellent applications are. But interviews with panel members have shown that they do not find it easy to operationalize the concept of excellence and its elements such as 'independence' or 'the ability to do groundbreaking research'. This results in uncertainty and in different approaches by different panelists. Panelists doubt about criteria deployed and express the need for clearer and operational criteria for 'excellence'.

Inspecting the review reports support this uncertainty. In some cases, a PI will get a low score because his/her "publications are not well cited", in others because "publications are not in the main multidisciplinary journals", and again in others because "the proposal has a high risk", despite that the ERC asks for "high risk – high gain" projects. Additionally, the workload of the panels is rather high, which often leads to heuristics-based decision making, and these heuristics may be stereotype-based [31].

**Research questions**

In this paper, we answer the following research questions: Do women compared to men *in the same field* and of *similar level of merit* (i) get a lower score from the panel, (ii) have a lower probability to proceed to the second evaluation step, and (iii) have a lower probability to receive the grant than men? And, (iv) if gender differences are varying between panels, what panel characteristics may explain gender bias?

**Measuring merit and gender stereotyping**

For selection of grant applications, *earlier contributions to science* are a core criterion, as reflected in e.g., the number of publications and their impact [14]. However, the bibliometric operationalization of merit is debated. For example, the validity of citations and productivity as quality criteria is strongly contested, as e.g., productivity would only measure quantity and not quality [59], but other research suggests that quantity and quality do go together [8], and are not opposing goals. Another criticism is that high productivity and impact lead to even higher productivity and impact (the Matthew effect), and therefore overestimate the contributions to science at the top end, and underestimate those at the bottom end. More fundamentally, indicators may be (gender) biased themselves. For example, women on average have more household tasks, and have lower academic positions than men. Consequently, women can invest less time in research and in building up an oeuvre, leading to lower bibliometric scores [6, 60]. This, however, does not invalidate the use of those performance indicators, but points at other instances of gender bias than in grant evaluation and decision.[10] In this case gender differences in private life, and in the university may lead to lower performance of women [63]. While acknowledging the relevance of several of these issues especially when using indicators at the individual level, using bibliometric indicators at group level can be considered adequate [64].

---

[10] Different forms of bias may add up, for example men may profit more from other forms of bias such as nepotism [61], and cognitive proximity [62]. But this can be modelled explicitly.

Bibliometric productivity and impact indicators are useful, but one should avoid a narrow approach to merit. Therefore, we also include other merit variables: the quality of the *collaboration network* of applicants, the number of *coauthors*, and previous *grants* the applicants did receive. Several other relevant merit-variables cannot be included here due to a lack of data: e.g., *prizes and awards*, and *community roles* such as editorships, board member of learned organization, program chair of conferences, and editorships of journals.

Gender stereotyping by panel members is expected to influence the evaluation of female applicants, as well as the grant decision. Furthermore, stereotypes are reflected in the language of the review reports. The literature shows that this can have different forms. (i) The same criteria may be used for men and women, but the evaluation of men tends to focus on the *presence* of required qualifications, whereas the evaluation of women tends to focus on the *absence* of qualifications [65]. One therefore would expect that men are evaluated in positive terms and women in negative terms. For example, independence is an important criterion, and may be used in the evaluation of men when they are independent, but not mentioned if they are not independent. For women it works in the opposite way [66]. (ii) Communication theory suggests that stereotyping generally leads to the use of negation words (like not, no, never) for the 'out group': negation bias [67]. This implies that also positive characteristics of women are formulated in negative terms ("She is not bad" versus "he is good".) [67]. This would add to the use of negation terms in reviews about women. (iii) Where the excellent scientist stereotype is male, male (agentic) traits (such as being assertive, confident, ambitious, dominant) are considered important [68]. One would therefore expect that the stronger the gender stereotypes in a panel, the more agentic terms are used in review reports, and in combination with the previous points different for men than for women. In review reports about men, one would expect agentic terms for traits of the applicants, whereas in review reports of women one would expect that the terms refer to the lack of those agentic traits. Gender stereotyping depends on implicit and automatic attitudes and opinions of panel members, but panels characteristics influence the role of stereotyping. For example, high work pressure may result in *decision heuristics* replacing the individual assessment of applicants [14]. And those heuristics may bring in gender stereotyping.

Apart from gender stereotyping, the panel composition and panel dynamics may also play a role [68a, 68b]. The latter we cannot test here, but the first we can: do the number of female panelists, the country-variety in the panel, and the level of experience of panel members have an effect on the level of bias?

## Data & methods

The data were collected from various sources: administrative data from the council, the application texts and the review texts, and WoS and Scopus for the bibliometric data.

### The variables

The dependent variables are the six *panel-scores* the applications get for (i) independent creative thinking; (ii) moving beyond the state of the art, (iii) having done groundbreaking research, (iv) the overall quality of the PI, and (v) the quality of the proposed project. In the second step there is also an assessment of (vi) the commitment of the PI with the project. The individual scores of panelists and external reviewers are on a 4-point scale. Two more dependent variables are the *decisions*: (vii) to proceed (1) or not (0) to the second step of the evaluation, and (viii) being granted or not (also 1/0).

The following independent variables are included in the analysis. Past performance is measured with 15 bibliometric indicators (Scival), based on the Scopus database and covering the period 1996-2014[11]. A factor analysis results in three performance dimensions: (i) absolute impact and output; (ii) relative (to the oeuvre) impact; (iii) journal impact. Also included are (iv) the number of previous grants, (v) the ranking of the host institution (as indicator of the quality of the collaboration network), and the (vi) the average number of coauthors.

Finally, several personal attributes are included: (vii) Age – being younger is often seen as positive, as one can argue that better researchers are at the same level as others at an earlier age; (viii) academic age (years since the PhD), as the earlier the PhD, the more time has been available for building a track record; (ix) research field – different fields may differ in the level of available resources; and (x) gender. As we use bibliometric data to measure past performance, we exclude those fields where journal articles are not the main publication type: the three panels covering most of the humanities, law, and most of the qualitative social sciences.

To analyze the *effect of panel characteristics* on the level of bias, we use the following variables: (i) gender stereotyping, measured by the frequency of negation words, and of agentic words in the review reports; (ii) the number (share) of women in the panel; (iii) the diversity of panels in terms of countries represented in the panel; (iv) the experience of the panel members (number of years in the panel); and (v) the work load (number of applicants per panelist). The LIWC tool [68c] was used to obtain the linguistic variables from the review reports. The resulting frequencies of the relevant linguistic categories have been averaged per panel for men and women separately.

---

[11] This covers the years that the applicants have been scientifically active at the moment of applying.

*Methods*

Using *mixed models* (SPSS 27) we predict the six panel scores for the PI and the project, with panels as random effects, and the merit variables and the personal characteristics as fixed effects. A significant regression coefficient for gender indicates gender bias in the scores.

*Logistic regression* is used to analyze the two binary decisions (proceeding to Step 2; being granted), with the same set of independent variables, and gender and the panel as factors – as the decisions are taken at the panel level. An analysis for men and women separately identified which regression coefficients differ, suggesting that the selection process differs for men and women. The relevant interaction (with gender) terms were added. For those variables that failed a linearity test, a squared variable is added to the model.

In order to inspect the gender effect in more detail, the predictive probabilities (PP) were calculated using STATA 16 (predicted margins). For logistic regression this is needed, as this provides the effect of gender for individual values of independent variables, something that cannot be done using the regression coefficient – which gives the gender effect only for one value (zero) of the independent variables [68d]. For each panel, the relative predicted probability: RPP = PPwomen/PPmen was calculated. Comparing this with the relative success rate RSR = SRwomen/SRmen gives a *gender bias indicator*: RPP/RSR. When this indicator is larger than one, there is gender bias against women. When the indicator is smaller than one, the panel shows gender bias in favor of women. And the higher the value of the indicator, the stronger the bias against women.

Finally, an exploratory analysis gives the *correlation* between the RPP/RSR (indicating gender bias) and several panel characteristics. This indicates what panel characteristics predict gender bias.

**Findings**

*Panel scores*

The five panel scores received in Step 1 correlate strongly. The overall score for the *project* and the four scores for the *PI* correlate between 0.78 and 0.85, suggesting that the evaluators hardly distinguish the quality of the PI and the quality of the proposed project. In Step 2, the correlations are lower but still strong. The five mixed models analyses of the scores in Step 1, and the six in Step 2, result in a consistent picture.

In Step 1, each of the scores is significantly positively influenced the by most of the independent variables: *absolute impact*, *journal impact*, the number of acquired *grants* and the *host score*. Only the *average number of coauthors* has a negative effect, suggesting that with many coauthors, the applicant's own contributions become less visible. The *relative impact* has a negative effect on the scores too[12]. Furthermore, we find that age influences the score negatively: the older the applicant, the lower the scores. Academic age has a positive effect on the scores. The latter is easily understood, as the more years (up to 7, or longer if the PI was granted an extension) after the PhD the grant application was submitted, the more time to build up an oeuvre and reputation (Table 1 for one of the dependent variables). Finally, after controlling for merit, women get uniformly lower scores than men (Table 2 - left).

**Table 1**: Panel score* by performance and gender

| Parameter | Estimate | Std. Error | df | t | Sig. | 95% CI - LB | 95% CI - UB |
|---|---|---|---|---|---|---|---|
| Intercept | -2.515 | 0.033 | 33.5 | -76.363 | 0 | -2.5822 | -2.4482 |
| Female | -0.081 | 0.022 | 2613.4 | -3.741 | 0 | -0.0385 | -0.1232 |
| Age | -0.066 | 0.010 | 2594.6 | -6.378 | 0 | -0.0867 | -0.0459 |
| Academic age | 0.079 | 0.011 | 2595.8 | 7.291 | 0 | 0.0574 | 0.0997 |
| Host score | 0.105 | 0.010 | 2594.7 | 10.386 | 0 | 0.0848 | 0.1243 |
| Total nr grants | 0.132 | 0.010 | 2594.6 | 13.218 | 0 | 0.1122 | 0.1513 |
| Average nr coauthors | -0.065 | 0.010 | 2594.6 | -6.371 | 0 | -0.0847 | -0.0448 |
| Total impact/output | 0.113 | 0.011 | 2595.1 | 10.517 | 0 | 0.0918 | 0.1338 |
| Journal impact | 0.202 | 0.012 | 2594.8 | 17.291 | 0 | 0.1789 | 0.2246 |
| Relative impact | -0.072 | 0.012 | 2594.7 | -6.047 | 0 | -0.0951 | -0.0485 |

\* Score for: proven ability for groundbreaking research.

In Step 2 the pattern is about the same, but not all variables have a significant effect, and some have hardly any effect (although the sign remains the same). One interesting deviation of the pattern is when predicting the commitment with the project: this is the only case where the number of other grants has a negative sign: The more grants, the significantly lower the score for commitment. This is also easily understood: with many grants, the PI has to divide his/her time over many projects. Also in Step 2, women receive lower scores than men.

---

[12] Relative impact correlates strongly negative (r = - 0.5) with absolute impact, suggesting that applicants with a high relative impact tend have a low productivity and a low total impact, which is in line with earlier findings [8].

Table 2 summarizes the gender effect: All coefficients are negative and in step 1 (Table 2, left part) also statistically significant, and in step 1 and 2 (right part), the Bs are about equal. In the second step the project score and the commitment score are both not significant, and groundbreaking research and the PI score are marginally significant. The conclusion is that after controlling for several merit variables, women get in both steps lower scores. Other studies suggest that the gender differences are mainly in the scores for the CV and not in the score for the project [50, 52]. In our analysis, the gender differences in the CV-score are about equal to the differences in the project score.

**Table 2.** Estimates of the effect of gender on review scores – (Mixed models)*

| | First step (N=2617) | | | | | Second step (N=708) | | | | |
|---|---|---|---|---|---|---|---|---|---|---|
| | B ** | SE | Sig | 95% CI | | B** | SE | Sig | 95% CI | |
| Groundbreaking research*** | -0.081 | 0.022 | 0.000 | -0.038 | -0.123 | -0.050 | 0.029 | 0.087 | 0.007 | -0.107 |
| Creative independent thinking | -0.065 | 0.010 | 0.000 | -0.084 | -0.045 | -0.062 | 0.029 | 0.033 | -0.005 | -0.118 |
| Beyond state of the art | -0.071 | 0.010 | 0.000 | -0.091 | -0.051 | -0.083 | 0.029 | 0.004 | -0.026 | -0.139 |
| PI score (CV) | -0.068 | 0.020 | 0.001 | 0.029 | 0.108 | -0.037 | 0.024 | 0.115 | -0.009 | 0.083 |
| Project score | -0.067 | 0.020 | 0.001 | 0.027 | 0.107 | -0.031 | 0.026 | 0.234 | -0.020 | 0.083 |
| Commitment to the project | | | | | | 0.003 | 0.028 | 0.908 | 0.058 | -0.052 |

* Controlling for merit.
** Female (1) versus male (0).
*** All independent variables are standardized at panel level

### The panel decisions

As described, the procedure consists of two consecutive binary decisions: (i) which proposals go to the second step, and (ii) which proposals are granted? Table 3 shows the results. Women have lower probability to go to Step 2 than men (Table 2, left column: B = -0.195 p = 0.132), after controlling for several covariates. This suggests that gender bias against women exists in the first step of the procedure. When making it into the second step, women have a higher probability to win the grant (Table 3, middle column: B=0.234, p=0.22). The overall analysis of who receives a grant shows no effect of gender anymore. The regression coefficients for gender are not statistically significant in the analyses of Step 1 and Step 2. That should not surprise as selection takes place at the panel level, and the bias may be different (against or in favor of women) in the different panels. We return to the panel level below.

The covariates have in most cases a similar effect in the three analyses. Absolute impact, journal impact, the host ranking, and the total number of grants have a positive effect, and relative impact, the number of coauthors, and age have a negative effect. In Step 2 (middle column), the sample has been reduced to about 25% excellent proposals, and most merit variables do not influence the decision anymore at this stage. Finally, academic age (years since the PhD) correlates positively with the predicted probability, and this effect is stronger for women than for men.

Table 3: Results of the logistic regression analyses of survival/success probability controlling for merit

| | From Step 1 to Step 2 | | | Grant – Step 2 | | | Grant – all applicants | | |
|---|---|---|---|---|---|---|---|---|---|
| | B | Sig. | Exp(B) | B | Sig. | Exp(B) | B | Sig. | Exp(B) |
| Constant | -1.261 | 0 | 0.283 | -0.744 | 0.051 | 0.475 | -2.552 | 0 | 0.078 |
| Panel | | 0.083 | | | 0.913 | | | 0.957 | |
| Female | -0.195 | 0.132 | 0.823 | 0.234 | 0.220 | 1.264 | 0.028 | 0.857 | 1.028 |
| Absolute impact/output | 0.189 | 0.022 | 1.208 | 0.179 | 0.042 | 1.196 | 0.165 | 0.011 | 1.18 |
| Journal impact | 0.566 | 0 | 1.762 | 0.093 | 0.413 | 1.098 | 0.589 | 0 | 1.802 |
| Relative impact | -0.361 | 0 | 0.697 | -0.027 | 0.765 | 0.973 | -0.236 | 0.001 | 0.789 |
| Mean number coauthors | -0.256 | 0 | 0.774 | -0.307 | 0.005 | 0.735 | -0.513 | 0 | 0.599 |
| Age | -0.193 | 0.002 | 0.824 | -0.173 | 0.096 | 0.841 | -0.284 | 0.001 | 0.753 |
| Academic age | -0.003 | 0.972 | 0.997 | 0.020 | 0.837 | 1.021 | 0.136 | 0.091 | 1.145 |
| Host ranking | 0.314 | 0 | 1.369 | 0.133 | 0.128 | 1.142 | 0.261 | 0 | 1.298 |
| Total nr grants | 1.257 | 0 | 3.517 | -0.067 | 0.329 | 0.935 | 0.752 | 0 | 2.122 |
| Absolute impact^2 | -0.041 | 0.082 | 0.96 | | | | | | |
| Total grants^2 | -0.181 | 0 | 0.835 | | | | | | |
| Journal impact^2 | | | | 0.109 | 0.030 | 1.115 | -0.108 | 0 | 0.897 |
| Relative impact * gender | 0.183 | 0.125 | 1.201 | | | | | | |
| Academic age * gender | 0.326 | 0.008 | 1.385 | | | | | | |
| Academic age * age | -0.102 | 0.055 | 0.903 | | | | | | |
| Mean coauthors * age | | | | | | | -0.229 | 0.03 | 0.795 |

All variables are normalized at the panel level.    Nagelkerke pseudo R^2 = .366    Nagelkerke pseudo R^2 = .087    Nagelkerke pseudo R^2 = .220

Inspecting the *marginal probability plots* for the decision in Step 1 (Figures 1 and 2) reveals that for four of the covariates, men have a higher predicted probability than women at the same score of the independent (merit) variable. For the others there is some interaction effect: for academic age, journal impact, and relative impact, top-performing women have a higher probability to proceed to Step 2 than men with equal scores. However, as the number of low-scoring applicants is much higher, numerically men have a higher probability to proceed to step 2, as the logistic regression indicated. A similar issue holds for the average number of coauthors: the less coauthors, the higher the predicted probability – and at that part of the distribution women fare better. For the applicants with high numbers of coauthors, it works in the opposite way: men do it better at that side of the distribution.

Similar graphs for the decision in step 2 and for the overall analysis are made (SI10), and as expected these show a small advantage for women and (Step 2) and no difference at all (overall analysis).

**Figure 1a**: Predicted probability to proceed to step 2 by bibliometric scores and gender

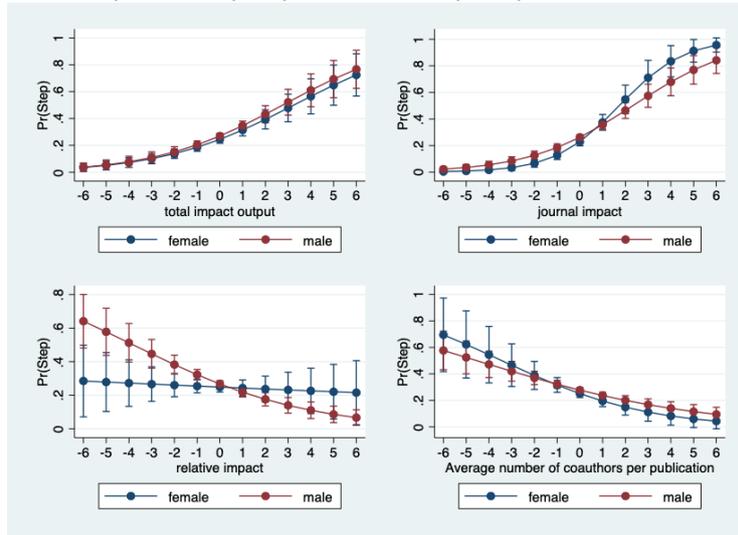

**Figure 1b**: Predicted probability to proceed to step 2 by several covariates and gender

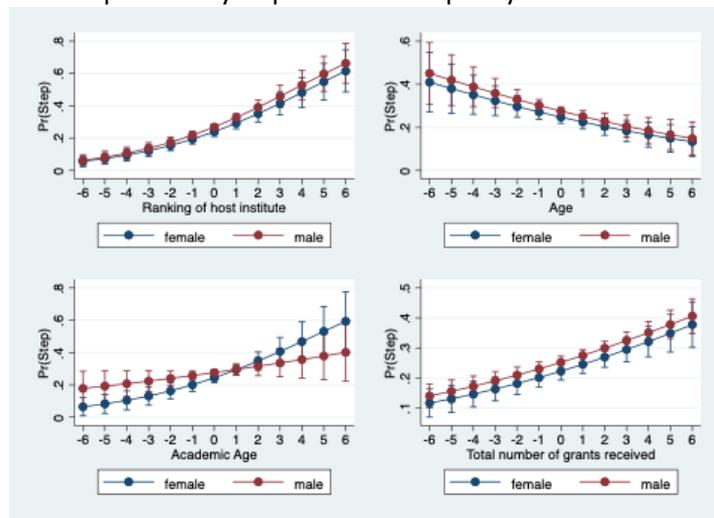

*Panel level analysis*

There may be panels with bias against women and panels with bias in favor of women. Bias may be *in different directions*, as can be measured by the 'gender bias indicator' defined above. Table 4 shows the results of the analysis at panel level. In the Life Sciences (LS) and the Social Sciences and Humanities (SH), there are more panels with bias against than in favor of women, and in the Physics and Engineering (PE) panels it is the opposite. Table 4 also shows the effects on the gender distribution of the grants. In LS, women should have received 3 more grants, and in SH 1.4 more. On the other hand, in PE women received 4 grants above expected. The numbers seem low, but as a share of all grants for women it is substantial: 7%, 16% and 11% in LS, PE and SH respectively.

**Table 4**. Locus and effects of gender bias at panel level  (N=22)

|  | All | LS | PE | SH |
|---|---|---|---|---|
| Panels with strong advantage for men | 4 | 2 | 1 | 1 |
| Panels with advantage for men | 2 | 1 |  | 1 |
| Neutral panels | 9 | 5 | 4 |  |
| Panels with advantage for women | 3 |  | 3 |  |
| Panels with strong advantage for women | 4 | 1 | 2 | 1 |
| grants gained (PE) and lost (LS, SH) by women | 0.2 | 3 | -4.2 | 1.4 |
| Share of all grants women received |  | 7% | -11% | 16% |

**Gender bias and panel characteristics**

Why do the scores for gender bias differ between the panels? One possible interpretation is that random variation in the decision-making process will lead to variety of bias scores at the panel level, without a systematic pattern. At panel level, men and women as a group sometimes win and sometimes lose, but over a longer period at the aggregated level, there may be equality. This would also explain the contradictory findings between studies on gender bias in grant allocation.

Another possibility is that the differences in gender bias relate to panel characteristics, such as panel composition, processes, procedures, and to the opinions and implicit gender stereotypes of the panelists. If there is no gender bias, one would expect no systematic correlation between panel characteristics and the bias indicator. The number of panels (22) in this study does not enable an in-depth multivariate analysis, but it allows for an exploratory descriptive analysis.

*Field differences.*

In fields with many female applicants (SH, LS), we found that bias tends to be against women, but in fields with low percentages of women (PE) it tends to be the opposite. This is similar to the findings of Van der Lee and Ellemers [50], who suggested that panelists in fields with many female researchers may become less aware that gender bias is still relevant, and (male and female) panelist may become much more critical towards women. On the other hand, in PE, we see that women have more success than expected. An explanation could be that within these fields a more diverse staff (and student population) is needed, and panels may therefore have some preference for women when deciding about grants[13]. We indeed found a positive relation between the share of women applicants in a field and the level of gender bias (r = 0.197). A consequence would be that if the share of women in PE-fields increases, the field may become more similar to LS and SH fields – leading to a stronger overall bias against women.

### First versus second step.

In the first step we found bias against women, and in the second step the bias was in favor of women. Both effects are not statistically significant, which is expected: The overall score is the aggregation of gender effects with opposite signs within individual panels. In Step 2, the panel has additional information: an interview with the applicants, a longer application text, and more review reports. However, the additional information does not lead to a more positive assessment of women, as in both Step 1 and Step 2, men get higher panel scores than women.[14] An alternative explanation could be related to the much higher work load in Step 1 than in Step 2. Small group research suggests that the higher *work load* in a group, the more members will use heuristics (such as gender stereotypes) for decision-making, instead of assessing individuals on their merit [30, 31]. If this would be the case here, one would expect a positive correlation between the workload and the bias indicator – but that was not the case (r = - 0.214).

### Gender stereotyping

One of the most discussed causes of gender bias is *gender stereotyping*, which can be identified through language use. *Negation* terms in review reports not only indicate negative evaluations of applicants but also reflect gender stereotypes about women [67, 69, 70]. As gender stereotyping may lead to gender bias against women [32, 71, 72], one would expect that the higher the gender bias score, the higher the amount of negation words used in review reports about women – which is indeed the case (r = 0.601, p = 0.003). The correlation between the gender bias indicator and negation words in review reports of men is lower (r =

---

[13] Research on selection of professors in these field suggests the same preference [68]

[14] In experiments with grant distributions by a lottery, the first step remains in panel review, as the lottery is implemented for a second step only (ref). Our findings suggest that lotteries may be detrimental to the interests of women, as a consequence of the bias in step 1. (We acknowledge Torger Möller for suggesting this implication of our findings).

0.350, p = 0.110), implying a difference: for men the use of negation words do not work negatively.

*Agentic characteristics* [73] are considered important for scientists. Men are expected to have agentic characteristics, but for women this is often seen as an exception. Therefore, one would expect that women are differently assessed in terms of agentic terms than men. We indeed find a positive relation between the bias indicator and the use of agentic terms in reviews about men (r = 0.273, p = 0.211), implying that the more agentic terms in the review, the higher bias *in favor* of men. For women we find an even stronger correlation between the gender bias indicator and the use of agentic terms (r = 0.458, p = 0.032). So the more agentic terms, the higher bias *against* women. When agentic characteristics are discussed, it is done in a positive way for men, but in a negative way for women.

### Panel composition

Despite the often-used policy of increasing the number of female panelists, we found that the higher the number of women panelists, the higher the gender bias score (r = 0.195), which also was found in other studies [47]. As gender equality is a visible priority within the ERC, one would expect that experienced reviewers (= not a first-year panel member) would be more aware of gender bias and act accordingly: a negative correlation between experience and gender bias. This works for men (r = -0.441, p = 0.04) but not for women (r = - 0.054). Experienced women seem to remain critical on women applicants, in line with the Queen Bee phenomenon [36]. Finally, diversity of a panel in terms of nationalities represented has a positive effect for women (r = -0.460, p = 0.031).

### Conclusions

First, controlling for merit, women receive lower scores by the panel than men, confirming the early results of Wennerås and Wold [37]. This conclusion is also in line with theories on how gender stereotypes would influence performance evaluation [32, 71, 72].

Second, bias in scores is not the same as bias in decisions. We found an overall bias in favor of men in Step 1 and an overall bias in favor of women in Step 2, but no bias in the final outcome. Furthermore, in 27% of the panels there is bias *against* women, and in 32% there is gender bias *in favor* of women, leading to an about equal overall result. Women lose in LS and SH, but win in PE.

Third, several panel characteristics do correlate with the level of gender bias, such as the share of female panelists, the share of experienced panelists, and the international variety of panel members. Using a linguistic approach to measure gender stereotyping, we found that the higher the level of gender stereotyping at the panel level, the stronger gender bias against women.

Fourth, several merit variables have a positive effect on the scores the applicants receive: having received the PhD at a young age, a high-ranking host institution, large (but not too large) number earlier grants, and a high level of absolute impact and output and a publishing in high impact journals. The *journal impact* has a stronger effect than the *total impact & output* variable. This may indicate that the review does not work optimal, because the journal impact indicator does measure the impact of the work only indirectly, as it is based on the journals where the papers were published in and not on the impact of the papers themselves.

Fifth, age correlates negatively with the scores and with success. Although this grant is meant for researchers with a low *academic age* (received a PhD in the last 7 years), the rules do not exclude older researchers. The age range of the applicants was from 28 to 66 years old, but not a single applicant over 46 received a grant. Clearly, age and not just academic age, influences decision making.

This study has several ***limitations:*** The findings at the panel level are explorative. Although the initial results are promising, a more robust analysis would need a larger sample of panels and funding instruments, in order to systematically relate the level of bias with the relevant characteristics of panels and panel members within their organizational and possibly national context. Another limitation relates to the merit variables, as several others could be included representing other merit dimensions, such as awards and prizes, and the applicant's independence [74]. Several lines of ***further research*** can be distinguished. First, women seem to apply less often than men, which leads to gender differences in winning research grants. Investigating application behavior may lead to interesting results. Second, several other forms of bias exist, based on nepotism, cognitive proximity, or ageism. These forms of bias may work out differently for men and women. Third, as the presentation of the proposal influences the selection process, the question comes up whether men and women differ in terms of self-presentation, and whether influences the selection process.

Summarizing, the study of the ERC Starting Grants leads to various conclusions that go beyond the specific case[15]. (i) Merit affects scores significantly, but (ii) gender does too: The scores women receive are lower than those of men after controlling for merit. (iii) Some panels show gender bias in favor of men, but also panels in favor of women. In this specific case study, the overall result is balanced. (iv) An important question is whether the variation at panel level is random or systematic. Our hypothesis, based on an explorative analysis, is that gender stereotyping (measured through word use in the review reports) strengthens gender bias against women, as does the share of female panel members. On the other hand, panel diversity in terms of nationalities reduces gender bias against women. This asks for further research.

---

[15] As the ERC has implemented new gender policies since our data collection, this study is not an evaluation of the current situation. However, this study has various practical implications, as these recent policies have not addressed several of the issues discussed in this paper: (i) The uncertainty about the role the various merit criteria should have in decision making; (ii) panel and field differences in criteria deployed, in the level of gender bias; (iii) the differences between the two steps concerning gender bias. The council introduced a training for panel members to handle implicit gender bias [75], which according to our analysis addresses an important problem. Whether this has been effective could be tested through a replication of this study with recent data.